\documentclass[lettersize,journal]{IEEEtran}
\usepackage{amsmath,amsfonts}
\usepackage{array}
\usepackage[caption=false,font=normalsize,labelfont=sf,textfont=sf]{subfig}
\usepackage{textcomp}
\usepackage{stfloats}
\usepackage{url}
\usepackage{verbatim}
\usepackage{graphicx}
\usepackage{lipsum}       
\usepackage{changepage}   
\usepackage{diagbox}
\def\BibTeX{{\rm B\kern-.05em{\sc i\kern-.025em b}\kern-.08em
    T\kern-.1667em\lower.7ex\hbox{E}\kern-.125emX}}
\usepackage{balance}
\usepackage[numbers,sort&compress]{natbib}
\usepackage{multirow}
\usepackage[flushleft]{threeparttable}
\usepackage{algorithm}
\usepackage[noend]{algpseudocode} 

\usepackage{makecell} 
\usepackage[colorlinks,urlcolor=black,citecolor=black,linkcolor=black,anchorcolor=black]{hyperref}
\newcommand{\PreserveBackslash}[1]{\let\temp=\\#1\let\\=\temp} 
\newcolumntype{C}[1]{>{\PreserveBackslash\centering}p{#1}}
\newcolumntype{R}[1]{>{\PreserveBackslash\raggedleft}p{#1}}
\newcolumntype{L}[1]{>{\PreserveBackslash\raggedright}p{#1}}
\let\OldStatex\Statex
\renewcommand{\Statex}[1][3]{%
  \setlength\@tempdima{\algorithmicindent}%
  \OldStatex\hskip\dimexpr#1\@tempdima\relax}

\begin{document}

\title{Solve Large-scale Unit Commitment Problems by Physics-informed Graph Learning} 

\author{Jingtao Qin,~\IEEEmembership{Student Member,~IEEE,} and Nanpeng Yu,~\IEEEmembership{Senior Member,~IEEE}
}



\maketitle

\begin{abstract}

Unit commitment (UC) problems are typically formulated as mixed-integer programs (MIP) and solved by the branch-and-bound (B\&B) scheme. The recent advances in graph neural networks (GNN) enable it to enhance the B\&B algorithm in modern MIP solvers by learning to dive and branch. Existing GNN models that tackle MIP problems are mostly constructed from mathematical formulation, which is computationally expensive when dealing with large-scale UC problems. In this paper, we propose a physics-informed hierarchical graph convolutional network (PI-GCN) for neural diving that leverages the underlying features of various components of power systems to find high-quality variable assignments. Furthermore, we adopt the MIP model-based graph convolutional network (MB-GCN) for neural branching to select the optimal variables for branching at each node of the B\&B tree. Finally, we integrate neural diving and neural branching into a modern MIP solver to establish a novel neural MIP solver designed for large-scale UC problems. Numeral studies show that PI-GCN has better performance and scalability than the baseline MB-GCN on neural diving. Moreover, the neural MIP solver yields the lowest operational cost and outperforms a modern MIP solver for all testing days after combining it with our proposed neural diving model and the baseline neural branching model.

\end{abstract}


\section*{Nomenclature}

\subsection{Superscripts and subscripts of variables}
\addcontentsline{toc}{section}{Nomenclature}
\begin{IEEEdescription}[\IEEEusemathlabelsep\IEEEsetlabelwidth{$V_1,V_2,V_3$}]
\item[$(\cdot)^n$] The $n$-th element of the vector.
\end{IEEEdescription}

\subsection{Sets, Vectors, and Constants}
\addcontentsline{toc}{section}{Nomenclature}
\begin{IEEEdescription}[\IEEEusemathlabelsep\IEEEsetlabelwidth{$V_1,V_2,V_3$}]
\item[$G$] Set of indices of generators.
\item[$T$] Set of indices of time periods.
\item[$N$] Number of buses.
\item[$M$] Number of transmission lines.
\item[$d_{i,t}$] Load demand of bus $i$ at time $t$.
\item[$R_t$] System spinning reserve demand at time $t$.
\item[$\overline{P}_g/\underline{P}_g$] Maximum/minimum power output of generator $g$.
\item[$SU_g/SD_g$] Start-up/shut-down ramp limit of generator $g$.
\item[$UT_g/DT_g$] Minimum up/down time limit of generator $g$.
\item[$RU_g/RD_g$] Ramp up/down time limit of generator $g$.
\item[$CU_g/ND_g$] Piece-wise start-up cost/delay vector of generator $g$.
\item[$PB_g/CB_g$] Piece-wise production amount/cost vector of generator $g$.
\item[$NL_g$] Number of segments of the piece-wise linear production cost curve of generator $g$.
\item[$F_l^+/F_l^-$] Maximum positive/negative capacity of transmission line $l$.
\item[$\alpha_{g,l}/\beta_{i,l}$] Power transfer distribution factor of generator $g$/bus $i$ to line $l$.
\end{IEEEdescription}

\subsection{Variables}
\addcontentsline{toc}{section}{Nomenclature}
\begin{IEEEdescription}[\IEEEusemathlabelsep\IEEEsetlabelwidth{$V_1,V_2,V_3$}]
\item[$\overline{p}$] Maximum available power output of generator $g$ at time $t$.
\item[$p'_{g,t}$] Power output above $\underline{P}_g$ of generator $g$ at time $t$.
\item[$x_{g,t}$] State transition variable (1 if $g$ remains operational at time $t$, 0 otherwise).
\item[$s_{g,t}/z_{g,t}$] Start up/shut down variable (1 if $g$ is turned on/off at time $t$, 0 otherwise).
\item[$f_{g,t}$] Start up cost of generator $g$ at time $t$.
\item[$v_{g,t}$] Production cost of generator $g$ at time $t$.
\end{IEEEdescription}


\section{Introduction}
\IEEEPARstart{U}{nit} Commitment (UC) is a fundamental optimization problem in the day-ahead electricity market. Given the supply offers, demand bids, and transmission system conditions, the Independent System Operator (ISO) needs to identify the optimal operation schedule of generators in a limited time. UC is most commonly formulated as a mixed-integer program (MIP) and solved by mathematical optimization algorithms including branch-and-cut (B\&C) algorithms \cite{gao2022internally}, Lagrangian relaxation (LR) \cite{ongsakul2004unit,hua2017representing}, Benders decomposition \cite{bertsimas2012adaptive,ramesh2021accelerated}, outer approximation \cite{yang2016multi}, ordinal optimization \cite{wu2013stochastic}, and column-and-constraint generation \cite{an2014exploring}. Among these approaches, B\&C is an effective method that combines branch-and-bound (B\&B) algorithms with cutting planes to improve the efficiency of the algorithm, which has been widely employed in modern solvers like Gurobi \cite{gurobi}, CPLEX \cite{cplex}, SCIP \cite{GleixnerEtal2018OO} and Xpress \cite{FICO}. For large-scale UC problems \cite{fu2013modeling}, the B\&B tree can be extremely large, resulting in a substantial computational time. 
\begin{figure*}[htbp]
\centerline{\includegraphics[width=14.8cm]{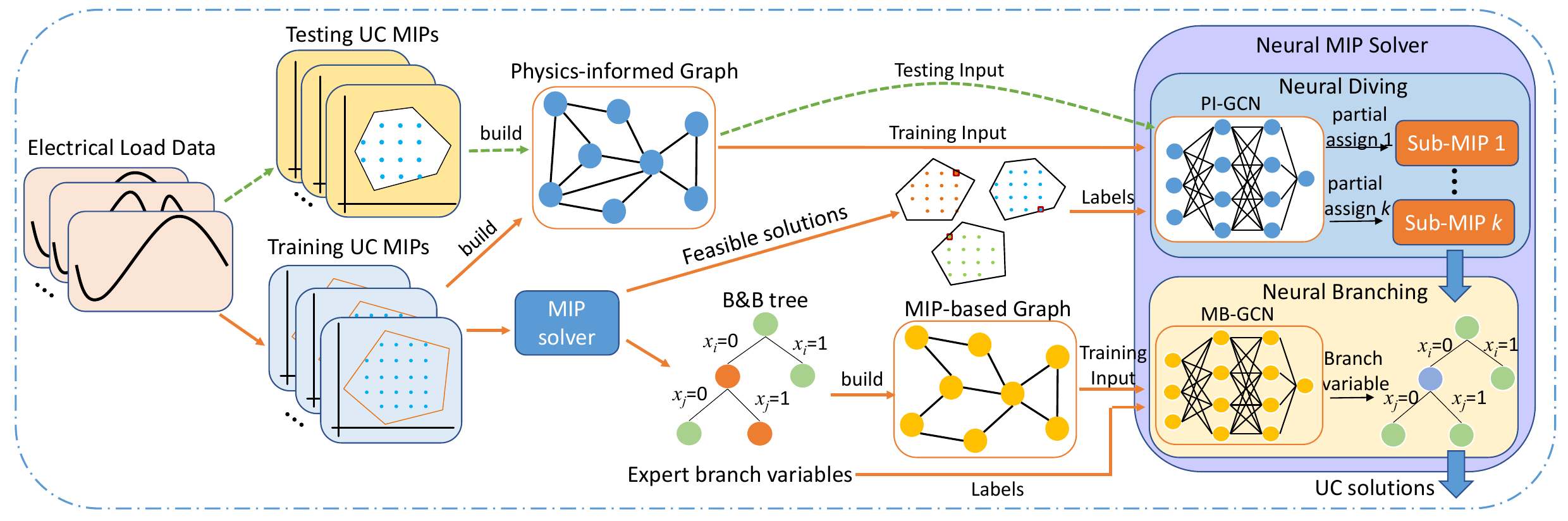}}
\caption{Overall framework of the proposed physics-informed graph learning to solve UC problems}
\label{fig:overall framework}
\end{figure*}

In practice, ISOs aim to achieve a lower optimality gap given a limited amount of computing time when solving UC problems. A lower optimality gap is desired because it indicates lower electricity market operation costs and higher market efficiency. Several strategies have been explored by researchers to improve the solution of large-scale UC problem models. One approach involves improving the tightness and compactness of the UC problem formulation, as demonstrated in previous studies \cite{morales2013tight,atakan2017state,yan2019systematic}. In addition, other decomposition and coordination approaches have been proposed for solving large-scale UC problems within strict time limits. For instance, the Surrogate Lagrangian Relaxation (SLR) technique has been utilized \cite{sun2018novel} to generate near-optimal solutions. Moreover, a novel quantum distributed model has been proposed \cite{nikmehr2022quantum} that employs decomposition and coordination to solve large-scale UC problems. 

Recent advancements in artificial intelligence have sparked a growing interest in applying machine learning (ML) techniques to enhance the performance of modern MIP solvers \cite{yang2021machine,chen2022security}. In the day-ahead electricity market, UC problems are solved on a daily basis with only minor changes in the input data. Most system characteristics, such as the parameters of generators or the topology of the transmission network, remain unchanged for the majority of the time \cite{xavier2021learning}. This presents an opportunity to use ML to learn from data generated by solving historical UC problems, thereby enhancing the performance of MIP solvers in tackling similar UC problems in the future. Many researchers have applied ML in modern MIP solvers to construct branching policies or obtain primal heuristics. In \cite{khalil2016learning}, E. B. Khalil, et al. use ML to construct a branching policy by collecting and learning features of candidate branching variables. Similarly, ML is utilized in \cite{alvarez2017machine} to create a fast approximation function from a set of observed strong branching decisions to build a branching policy. In \cite{marcos2015machine}, ML is used to create a function to estimate the number of nodes of sub-problems to parallelize B\&B. Additionally, E. B. Khalil, et al. use ML to predict the nodes where heuristics should be executed in \cite{khalil2017learning}. E. B. Khalil also uses reinforcement learning (RL) to learn heuristic algorithms over graphs in a subsequent work \cite{khalil2017alearning}. 

Graph Neural Networks (GNN) have been applied to address many problems in power systems \cite{liao2021review}. For example, the application of GNN for learning to branch in B\&B algorithms has been demonstrated to effectively reduce the running time of MIP \cite{gasse2019exact,gupta2020hybrid,labassi2022learning}. DeepMind \cite{nair2020solving} develops a GNN model proposed in \cite{gasse2019exact} which is built from the MIP model and applies it to neural diving and neural branching for solving problems including electric grid optimization \cite{knueven2020mixed}. However, when dealing with large-scale UC problems, the graph built from the MIP model is extremely complex, thus making the algorithm computationally expensive. To improve the performance and scalability of GNN models in the realm of neural diving for solving UC problems, we introduce a novel physics-informed graph convolutional network (PI-GCN) that seeks to leverage the inherent structure and relationships present in physical power systems for neural diving. Then, we adopt the MIP model-based graph convolutional network (MB-GCN) proposed in \cite{gasse2019exact} for neural branching and integrate it with neural diving to build a neural MIP solver. 

As illustrated in the overall framework in Fig. \ref{fig:overall framework}, PI-GCN is trained to find high-quality variable assignments for neural diving, while MB-GCN is trained to select the ideal variables for branching within the B\&B tree for neural branching. The distinctive aspect of our proposed PI-GCN lies in its incorporation of physical information derived from UC problems, including load demands, generator characteristics, and transmission line capacities. Furthermore, we leverage feasible solutions obtained from the MIP solver to create highly accurate training labels, enhancing the model's ability to generalize and adapt effectively to unseen problems. As for neural branching, MIP model-based graphs are generated from the MIP solver and utilized as training inputs, while branch variables selected by expert branch policy serve as training labels.

In contrast to previous studies, the unique contributions of this paper are as follows:
\begin{adjustwidth}{10pt}{}
1) To the best of our knowledge, this is the first study to integrate GNN with a physical power system to tackle large-scale UC problems. Our proposed PI-GCN model for neural driving surpasses the baseline MB-GCN, as it not only yields a solution with lower operational cost but also reduces the reliance on computing resources by significantly shrinking the graph size at the same time.

2) To exploit the underlying features of various components of power systems, we construct a spatiotemporal graph using time series load data information, and a spatial graph using the information on generators and transmission lines. Additionally, we propose a hierarchical graph convolution model comprising a spatiotemporal graph convolution module, an edge-conditioned graph convolution module, and a variable map module to handle diverse graph types and to map the physical input to the variable output.

3) We integrate the neural diving and neural branching with a modern MIP solver to establish a neural MIP solver that is tailored for UC problems, which exhibits superior performance than the commercial MIP solver in terms of achieving a lower operational cost and possesses exceptional generalization capabilities for solving previously unseen UC problems.
\end{adjustwidth}

The remainder of the paper is organized as follows: the preliminaries are given in Section II; the proposed PI-GCN model for neural diving is introduced in Section III; the baseline MB-GCN model for neural diving and branching is present in Section IV; numerical studies are given to illustrate the performance of our proposed model in Section V, and conclusions are drawn in Section VI.

\section{Preliminaries} 

In this section, the formulation of UC problems is first presented. Then, we review the concept of the branch-and-bound tree. Finally, the basics of diving and strong branching policy are covered.

\subsection{UC Formulation} \label{UC model}
In this paper we adopt the state transition model proposed in \cite{atakan2017state} to formulate UC problems. As shown in (\ref{eq:obj}), the objective is to find the optimal operation schedule of units, which yields the minimum overall operational cost. Here, we consider six types of constraints: namely, the load/spinning reserve demand constraints (\ref{eq:load})-(\ref{eq:spinning}), generator output constraints (\ref{eq:generation1})-(\ref{eq:generation2}), minimum up and down time limits (\ref{eq:minimum up})-(\ref{eq:minimum dn}), ramping limits (\ref{eq:ramp1})-(\ref{eq:ramp2}), transmission line capacity constraints (\ref{eq:line constraint}) and  the state-transition constraints (\ref{eq:state-transition})-(\ref{eq:state-transition-3}).

\begin{equation} \label{eq:obj}
    \min \sum_{g\in G} \sum_{t\in T} \{f_{g,t} + v_{g,t} + CU_{g}^1 s_{g,t}+CB_g^1(s_{g,t}+x_{g,t})\}
\end{equation}
\begin{gather}
    s.t. \sum_{g\in G} \left(p'_{g,t} + \underline{P}_g(s_{g,t}+x_{g,t}) \right) \geq \sum_{i=1}^{N} d_{i,t}, \forall t \in T \label{eq:load} \\
    \sum_{g\in G} \overline{p}_{g,t} \geq \sum_{i=1}^{N} d_{i,t} + R_t, \forall t \in T \label{eq:spinning}\\
    \overline{p}_{g,t} \geq p'_{g,t} + \underline{P}_g(s_{g,t}+x_{g,t}),\forall g \in G, \forall t \in T \label{eq:generation1} \\
    \overline{p}_{g,t} \leq \overline{P}_g(s_{g,t}+x_{g,t}) + (SD_g-\overline{P}_g)z_{g,t+1}, \forall g \in G, \forall t \in T \label{eq:generation2} \\
    \sum_{i=t-UT_g+1}^{t-1} s_{g,i} \leq x_{g,t}, \forall g \in G, \forall t \in T \label{eq:minimum up}\\
    \sum_{i=t-DT_g}^t s_{g,i} \leq 1-x_{g,t-DT_g}, \forall g \in G, \forall t \in T \label{eq:minimum dn}\\
    \overline{p}_{g,t}-p'_{g,t-1} \leq SU_g s_{g,t} + (RU_g + \underline{P}_g)x_{g,t},\forall g \in G, \forall t \in T \label{eq:ramp1}\\
    p'_{g,t-1}-p'_{g,t} \leq (SD_g-\underline{P}_g)z_{g,t}+RD_g x_{g,t}, \forall g \in G, \forall t \in T \label{eq:ramp2} \\
    F_m^- \leq \sum_{g=1}^{|G|} \left(p'_{g,t}+\underline{P}_g(x_{g,t}+s_{g,t}) \right) \alpha_{g,l} -\sum_{i=1}^{N} d_{i,t} \beta_{i,l} \leq F_m^+, \notag \\ \forall m \in 1,\dots,M, \forall t\in T \label{eq:line constraint} \\
    s_{g,t-1} + x_{g,t-1} = z_{g,t} + x_{g,t}, \forall g \in G, \forall t \in T \label{eq:state-transition}\\
    f_{g,t} \geq (CU_{g}^n-CU_{g}^1)\left(s_{g,t}-\sum_{i=DT_g}^{ND_{g}^n} s_{g,t-i} - x_{g,t-ND_{g}^n} \right),  \notag \\ \forall g \in G, \forall t \in T, \forall n \in {1,\dots,|ND_g|} \label{eq:state-transition-2} 
\end{gather}
\begin{gather}
    v_{g,t} \geq vc_g^n (p'_{g,t}+\underline{P}_g-PB^n_{g})+CB_{g}^n-CB_{g}^1, \notag \\
    \forall g \in G, \forall t \in T, \forall n \in 1,\dots, NL_g \label{eq:state-transition-3}
\end{gather}

\subsection{Branch-and-bound Algorithm}
The UC problem formulated in subsection \ref{UC model} is a mix-integer linear program (MIP), which has the general form in (\ref{eq:MILP}):
\begin{equation}\label{eq:MILP}
    \underset{\mathbf{x}}{\text{argmin}} \{\mathbf{c}^T\mathbf{x}|\mathbf{Ax} \leq \mathbf{b}, \mathbf{l} \leq \mathbf{x} \leq \mathbf{u}\},
\end{equation}
where $\mathbf{x} \in \mathbb{Z}^p \times \mathbb{R}^{n-p}$ is the decision variable vector, $\mathbf{c} \in \mathbb{R}^n$ is the coefficient vector of the objective function, $\mathbf{A} \in \mathbb{R}^{m\times n}$ is the coefficient matrix of the constraints, $\mathbf{b} \in \mathbb{R}^m$ is the right-hand-side vector of the constraints, $\mathbf{l}, \mathbf{u} \in \mathbb{R}^n$ are the lower and upper bound vectors for the decision variables, and $p \leq n$ is the dimension of integer variables.

Branch-and-bound (B\&B) is the most commonly used technique to solve MIP. By relaxing the integrality constraint, we can obtain a linear program (LP) whose solution gives a lower bound to (\ref{eq:MILP}). If the solution of the relaxed LP does not satisfy the integrality requirement, we then decompose the relaxed LP into two sub-problems whose solutions are feasible solutions to (\ref{eq:MILP}) by selecting a variable that does not respect integrality and splitting the feasible region according to (\ref{eq:splitting}):
\begin{equation} \label{eq:splitting}
    x_i \leq \lfloor x^*_i \rfloor \vee x_i \geq \lceil x^*_i \rceil, i \leq p |x^*_i \notin \mathbb{Z},
\end{equation}
where $\lfloor \cdot \rfloor,\lceil \cdot \rceil$ are the floor and ceiling functions. 

By performing this binary decomposition iteratively, the B\&B algorithm generates a search tree where the best LP solution gives a lower bound and the best integral LP solution (if any) gives an upper bound. The solving process stops when the gap between the upper bound and lower bound is small enough or when search limits based on time or visited nodes are reached. 
\subsection{Diving and Strong Branching}

Diving is a set of primal heuristics of exploring the branch-and-bound tree in a depth-first manner. It involves sequentially fixing integer variables until either a leaf node is reached or the LP problem becomes infeasible. Diving is often used as a means of efficiently searching for feasible solutions to complex optimization problems. Unlike traditional diving techniques, such as those described in \cite{berthold2006primal,eckstein2007pivot}, which begin at any node and continue descending until a leaf node is reached, the diving approach in \cite{nair2020solving} and this paper is performed only from the root node. Instead of fully descending the tree, we only partially explore the tree and then use an MIP solver to solve the remaining sub-MIPs. 

The method to select a variable for branching is essential for the branch-and-bound algorithm. Typically, the measurement for the quality of branching on a variable is the improvement of the dual bound \cite{khalil2016learning}. Given a node $N$ with LP value $\hat{z}$, LP solution $\hat{x}$, and candidate variable set $\mathcal{C}$, two children nodes, $N^-_i$ and $N^+_i$, are created by branching on variable $x_i$ in a downwards and upwards direction, respectively. These children nodes have LP values of $\hat{z}^-_i$ and $\hat{z}^+_i$, which are only considered if they are feasible. In the case that $N^{-}_{i} (N^{+}_{i})$ is infeasible, $\hat{z}^{-}_i (\hat{z}^+_i)$ is assigned a very large value. Then, the strong branching (SB) score of branching on variable $x_i$ can be calculated as in \eqref{eq:SB score}:
\begin{equation}\label{eq:SB score}
    SB_i = \max{\{\hat{z}^-_i-\hat{z},\epsilon\}} \times \max{\{\hat{z}^+_i-\hat{z},\epsilon\}},
\end{equation}
where $\epsilon$ is a small constant (e.g. $10^{-6}$).

\section{Physics-informed Graph Convolutional Model for Neural Diving}
In this section, we present our proposed physics-informed graph convolutional network (PI-GCN) for neural diving, which enables us to fix certain integer variables when solving large-scale UC problems. We first give the definition of the physical network information-based graphs. Then, we present the network architecture of the proposed PI-GCN. Next, we introduce the loss function for neural diving. Finally, we propose a parallel sub-MIPs solving framework.

\subsection{Graph Definition for UC Problems} \label{sec:proposed graph definition}
\begin{figure}[htbp]
\centerline{\includegraphics[width=7.5cm]{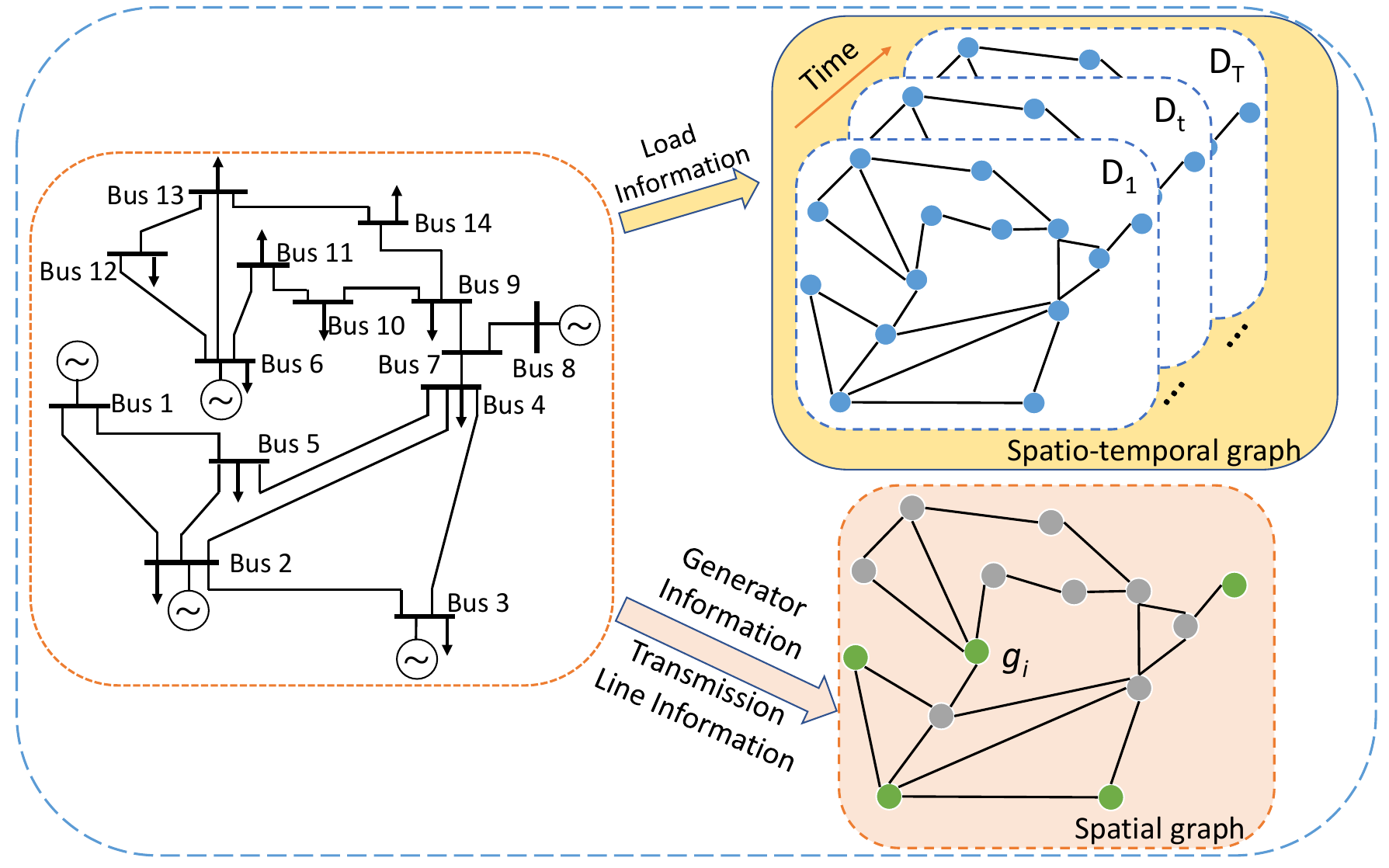}}
\caption{Physical network information-based graphs}
\label{fig:physical graph}
\end{figure}

As shown in Fig. \ref{fig:physical graph}, the physical network for UC problems can be deconstructed into two distinct sub-graphs, namely the spatiotemporal graph and the spatial graph, to capture and represent different aspects of the power system, and gain flexibility and adaptability to the dual challenges of temporal variations and spatial dependencies inherent in UC problems.
\subsubsection{Spatiotemporal Graph}The spatiotemporal graph is defined as $\mathcal{G}_t=(\boldsymbol{D}_t,\mathcal{E},\mathcal{A})$, where $\boldsymbol{D}_t$ is the set of nodes on time step $t$, corresponding to the load information at time $t$, $|\boldsymbol{D}_t|=N$. $\mathcal{E}$ is the set of edges, $|\mathcal{E}|=M$. $\mathcal{A}$ is the adjacent matrix of $\mathcal{G}_t$, $\mathcal{A}\in \{0,1\} ^{N \times N}$. $\mathcal{A}_{ij}$ equals to 1 if node $i$ and node $j$ are connected, and 0 otherwise. The feature of node $i$ is defined as the load data of bus $i$ at time step $t$.

\subsubsection{Spatial Graph}The spatial graph is defined as $\mathcal{G}=(\boldsymbol{G},\boldsymbol{E},\mathcal{A})$, where $\boldsymbol{G}$ is the set of nodes corresponding to the generator information, $|\boldsymbol{G}|=N$. $\boldsymbol{E}$ is the set of edges that correspond to transmission line information, $|\boldsymbol{E}|=M$. The feature of node $i$ is defined as $\boldsymbol{g}_i=[\boldsymbol{PB}_i,\boldsymbol{CB}_i,RU_i,RD_i,SU_i,SD_i,v^0_i,p^0_i]$, where $v^0_i$ is the initial on/off status of generator $i$, and $p^0_i$ is the initial output of generator $i$. Note that if node $i$ does not have a generator, then $\boldsymbol{g}_i$ will be a zero vector. The feature of edge $m$ is defined as $\boldsymbol{e}_m=[F^+_m,F^-_m,R_m,S_m]$, where $R_m$ and $S_m$ are the reactance and susceptance of line $m$, respectively. 

\vspace{-0.2cm}

\subsection{Network Architecture}
\begin{figure}[htbp]
\centerline{\includegraphics[width=8.5cm]{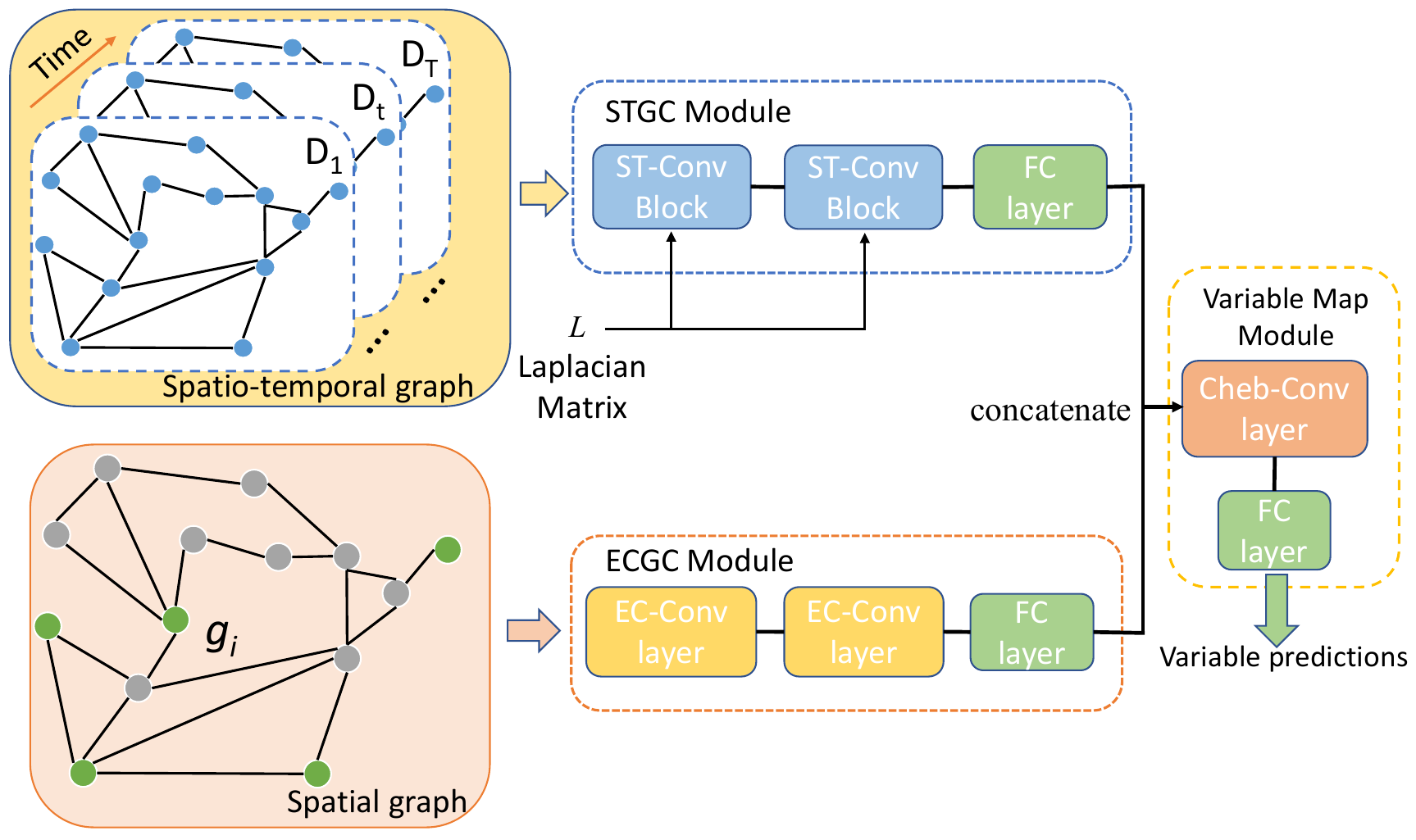}}
\caption{Network architecture of PI-GCN}
\label{fig:proposed network}
\end{figure}

The architecture of the PI-GCN designed for neural diving is illustrated in Fig. \ref{fig:proposed network}. The data processing begins by routing the spatiotemporal graph through the spatiotemporal graph convolution (STGC) module, while simultaneously passing the spatial graph through the edge-conditioned graph convolution (ECGC) module. Subsequently, the outputs of these two modules are combined into a unified graph, which is then fed into the variable map module. The specific functionalities of each module are elaborated upon in the following subsections.

\subsubsection{STGC Module}
The STGC module is effective in discerning patterns and dependencies over time at different nodes, which consists of two spatiotemporal convolution (ST-Conv) blocks and a fully connected (FC) layer. Each ST-Conv block contains two temporal gated convolution layers and one spatial graph convolution layer in the middle \cite{yu2017spatio}.

The temporal gated convolution layer consists of a 1-D causal convolution with a $K_t$ width kernel followed by a Gated Linear Unit (GLU) as a non-linearity as shown in (\ref{eq:GLU}). For every node within the graph $\mathcal{G}$, the temporal convolution examines $K_t$ neighboring points in the input time series load data without employing padding, resulting in a reduction of sequence length by $K_t-1$ on each iteration:
\begin{equation}\label{eq:GLU}
    \mathbf{D}' = (\mathbf{D}*\mathbf{W}) \odot \sigma(\mathbf{D}*\mathbf{V}),
\end{equation}
where $\mathbf{D}\in \mathbb{R}^{N\times T\times 1}$ is the input node features of the spatiotemporal graph, $\mathbf{W},\mathbf{V} \in \mathbb{R}^{K_t\times 1 \times C_o}$ are the convolution kernels with the same size, $C_o$ is the size of output channels, and $\sigma(\cdot)$ is the sigmoid gate function. 

We use the Chebyshev spectral convolution network \cite{defferrard2016convolutional} as the spatial graph convolution layer. The operator equation is given by (\ref{eq:chebyshev}):
\begin{equation}\label{eq:chebyshev}
    \mathbf{X}'_t=\sum_{k=1}^{K_s} \mathbf{Z}^{(k)} \cdot \mathbf{W}^{(k)},\ t\in T-K_t+1,
\end{equation}
where $K_s$ is the graph convolution kernel size, $\mathbf{W}^{k}$ is the $k$-th weight matrix, $\mathbf{Z}^{(k)}$ is $k$-th Chebyshev polynomial which is computed by (\ref{eq:cheby_term}) with $\mathbf{Z}^{(1)}=\mathbf{D}'_t, \mathbf{Z}^{(2)}=\hat{\mathbf{L}}$.
\begin{equation}\label{eq:cheby_term}
\begin{split}
    \mathbf{Z}^{(k)} &= 2\hat{\mathbf{L}}\cdot\mathbf{Z}^{(k-1)}-\mathbf{Z}^{(k-2)}.
\end{split}
\end{equation}
Here $\mathbf{D}'_t$ represents the node embedding matrix at time $t$, which is the output generated by the temporal gated convolution layer. $\mathbf{\hat{L}}$ is the scaled and normalized Laplacian matrix of the graph, $\mathbf{\hat{L}}=2\mathbf{L}/\lambda_{\max}-\mathbf{I}$. $\lambda_{\max}$ is the maximum eigenvalue of $\mathbf{L}$. $\mathbf{L}$ is the Laplacian matrix of the graph, and $\mathbf{L}=\mathbf{I}-\mathcal{D}^{-\frac{1}{2}}\mathcal{A}\mathcal{D}^{-\frac{1}{2}}$. $\mathcal{D}$ is the degree matrix of the graph. By performing $K_s$-localized convolutions using polynomial approximation, the information from nodes within a $K_s$ neighborhood will be propagated to the central node.

\subsubsection{ECGC Module}

The ECGC module can effectively incorporate spatial dependencies and interpret how nodes interact based on the characteristics of their connecting edges, which consists of two edge-conditioned convolutional (EC-Conv) layers that can manage graphs with edge features and an FC layer. 

For the EC-Conv layer, $g_i$ is propagated to $g'_i$ using the operator shown in (\ref{eq:ECC}):
\begin{equation}\label{eq:ECC}
    \boldsymbol{g}'_i = \mathbf{W}\boldsymbol{g}_i + \sum_{j \in \mathcal{N}(i)}\boldsymbol{g}_j \cdot h_\Theta(\boldsymbol{e}_{i,j}),
\end{equation}
where $\mathbf{W}$ is a weight matrix, $\mathcal{N}(i)$ is the set of the neighbor nodes of node $i$, $h_\Theta$ is a neural network to propagate edge features, i.e. a multi-layer perceptron, and $\boldsymbol{e}_{i,j}$ is the edge feature vector between node $i$ and node $j$.

\subsubsection{Variable Mapping Module}
In order to transform the processed graph-based information into a variable-centric representation, we introduce a novel variable mapping module. The variable mapping module consists of a Chebyshev convolution layer and an FC layer. In order to map the embeddings of generator nodes to decision variables, we first apply a Chebyshev convolution layer to propagate the embeddings of all nodes to generator nodes. As aforementioned, embeddings of the first-order neighbor nodes are propagated to their central node when the graph convolution kernel size $K_s=1$. Similarly, embeddings of the first order, as well as the second order neighbor nodes, are propagated to their central node when $K_s=2$. 

Then, we filter the generator nodes and pass them to an FC layer. At last, the output of the FC layer will be reshaped to match the dimensions of the decision variables.

\begin{figure}[htbp]
\centerline{\includegraphics[width=8.5cm]{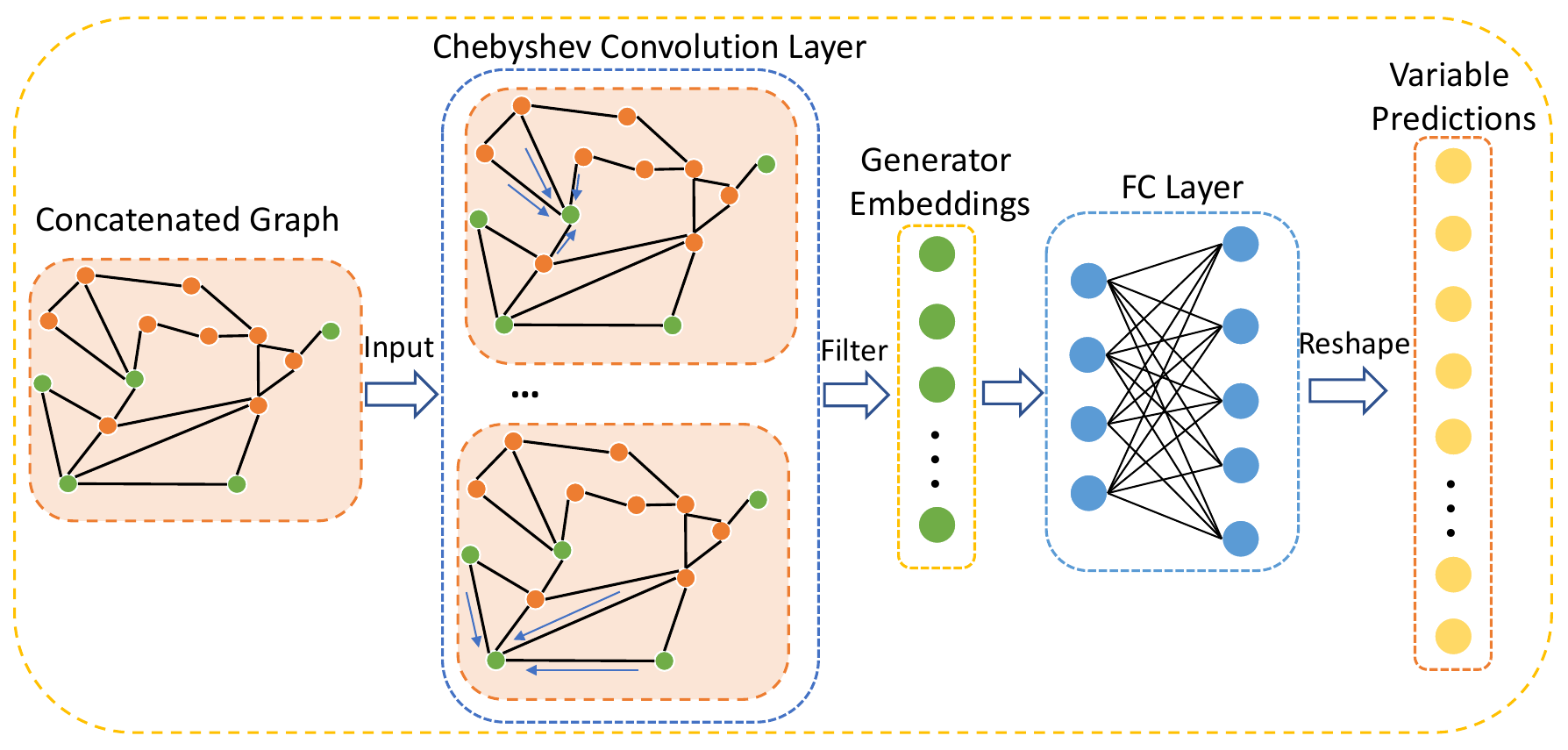}}
\caption{Variable mapping module}
\label{fig:variable map}
\end{figure}

\vspace{-0.5cm}
\subsection{Loss Function for Neural Diving}
Assuming that $\mathcal{I}$ represents the set of MIP instances generated under different load profiles in section \ref{UC model}, and $\mathbb{G}$ is the mapping function that converts a certain instance $I$ into a graph $\mathcal{G}$. In this paper, we propose to only predict the state variable $x_{g,t}$ in the state-transition model. $\boldsymbol{x}$ is a vector containing the state variables $x_{g,t}$ of instance $I$, and $C(x)$ is the objective function of $I$. The loss function for neural diving is given in (\ref{eq:diving loss}):
\begin{equation} \label{eq:diving loss}
    L(\theta)=-\frac{1}{N}\sum_{i=1}^N \exp{\left(-\beta C_i(\boldsymbol{x}^*_i)\right)}\log p_\theta(\boldsymbol{x}_i|\mathbb{G}(\mathcal{I}_i)),
\end{equation}
where $N$ is the number of instances, $\boldsymbol{x}^*_i$ is the best solution to instance $\mathcal{I}_i$ obtained by MIP solver, and $\beta$ is the weight used to scale the objective value.

We assume that all binary variables are independent and define the parameterized conditionally-independent probability distribution $p_\theta(\boldsymbol{x}|\mathcal{G})$ for graph $\mathcal{G}=\mathbb{G}(I)$ in (\ref{eq:pro distri}):
\begin{equation} \label{eq:pro distri}p_\theta(\boldsymbol{x}|\mathcal{G})=\prod_{d\in\mathcal{D}}p_\theta(\boldsymbol{x}_d|\mathcal{G}),
\end{equation}
where $\mathcal{D}$ is the set of dimensions of $x$. For each dimension of $x$, the parameterized probability is calculated by (\ref{eq:pro}):
\begin{equation} \label{eq:pro}
p_\theta(\boldsymbol{x}_d|\mathcal{G})=\left\{
\begin{aligned}
&\frac{1}{1+\exp{(-\boldsymbol{y}_d)}} \ \text{if} \ x_d=1,\\ 
&\frac{\exp{(-\boldsymbol{y}_d)}}{1+\exp{(-\boldsymbol{y}_d)}} \ \text{if} \ x_d=0,
\end{aligned}
\right.
\end{equation}
where $\boldsymbol{y}_d$ is the output of the graph convolution network corresponding to dimension $d$.

\vspace{-0.3cm}
\subsection{Parallel Sub-MIPs Solving}
After training the graph convolution network to infer state variables, we employ the obtained probability distribution $p_\theta(x|\mathcal{G})$ to test various instances. 
\begin{algorithm}[htbp]
    \caption{Parallel Sub-MIPs Solving}
    \label{algo:parallel}
     Input: learned distribution $p_\theta$, test instance $I$, accuracy threshold $\epsilon$, sample ratio set $\mathcal{P}$, \\
     Output: minimum objective value $c_{\min}$ and solution $\boldsymbol{x}_{\min}$ \\
     Candidate variable set: $\Psi:= \{\}$ \\
     Sub-MIPs set $\hat{\mathcal{I}}:=\{\}$ \\
     Solutions set $X:= \{\}$ \\
     Objective values set $C:= \{\}$ 
    \begin{algorithmic}[1]
    \For{$d=1,\dots,D$}{}
        \If{$Acc(\boldsymbol{x}_d)\geq\epsilon$}{}
        \State Append $\boldsymbol{x}_d$ to $\Psi$
        \EndIf
    \EndFor
    \State $\hat{\Psi}=\text{sort}(\Psi,\text{descending=true})$
    \For{$\rho \in \mathcal{P}$}{} 
        \For{$\boldsymbol{x}_d \in \hat{\Psi}\left[0:\lfloor\rho\cdot||\hat{\Psi}||\rfloor\right]$}{}
            \State Add constraint $\boldsymbol{x}_d=\lfloor p_\theta(\boldsymbol{x}_d|\mathbb{G}(I))+\frac{1}{2} \rfloor$ to $I$
        \EndFor
        \State Append obtained sub-MIP $\hat{I}$ to $\hat{\mathcal{I}}$.
    \EndFor
    \For{thread $i=1,\dots ||\mathcal{P}||$}{}
        \State Assign sub-MIP $\hat{\mathcal{I}}_i$ to thread $i$.
        \If{$\hat{\mathcal{I}}_i$ is feasible}{}
            \State solve sub-MIP $\hat{\mathcal{I}}_i$ and obtain $\boldsymbol{x}^*_i, c_i$
        \EndIf
        \State Append $\boldsymbol{x}^*_i$ to $X$ and $c_i$ to $C$
    \EndFor
    \State $n=\text{argmin} \ C$
    \State $c_{\min}=C[n],\boldsymbol{x}_{\min}=X[n]$
    \end{algorithmic}
\end{algorithm}

The process involves selecting specific binary variables, assigning them values of 0 or 1, and subsequently solving the remaining sub-MIP with an MIP solver. The detailed procedure is outlined in Algorithm \ref{algo:parallel}. To start, we compute the accuracy of each binary decision variable on the validation dataset using (\ref{eq:acc}):
\begin{equation} \label{eq:acc}
    Acc(\boldsymbol{x}_d)=\frac{1}{N}\sum_{i=1}^N\mathbf{1}\left(\lfloor p_\theta(\boldsymbol{x}_d|\mathbb{G}(\mathcal{I}_i))+\frac{1}{2} \rfloor=\boldsymbol{x}^*_{i,d}\right),
\end{equation}
where N is the number of instances in the validation dataset, $x^*_{i,d}$ is the true value of the solution to instance $\mathcal{I}_i$ corresponding to dimension $d$.

Next, we proceed by selecting binary variables with accuracy exceeding a specified threshold and arranging them in descending order, forming a candidate variable set. Subsequently, we pick specific variables for various sample ratios, setting their values to 0 or 1 according to their respective probability distribution. These chosen variables are then integrated into the initial MIP instance, creating sub-MIPs. Finally, we solve these sub-MIPs in parallel and identify the best solution that yields the minimum objective value for the original MIP.

\section{MIP model-based Graph Convolutional Model for Neural Diving and Branching} 

In this section, the baseline MIP model-based graph convolutional model (MB-GCN) for neural diving and branching technique is presented. We first introduce the MIP model-based graph. Then, we briefly review the network architecture of the baseline graph convolution model \cite{gasse2019exact}. Finally, we give the loss function for neural diving and neural branching.

\subsection{MIP model-based Graph Definition}
\begin{figure}[htbp]
\centerline{\includegraphics[width=8.0cm]{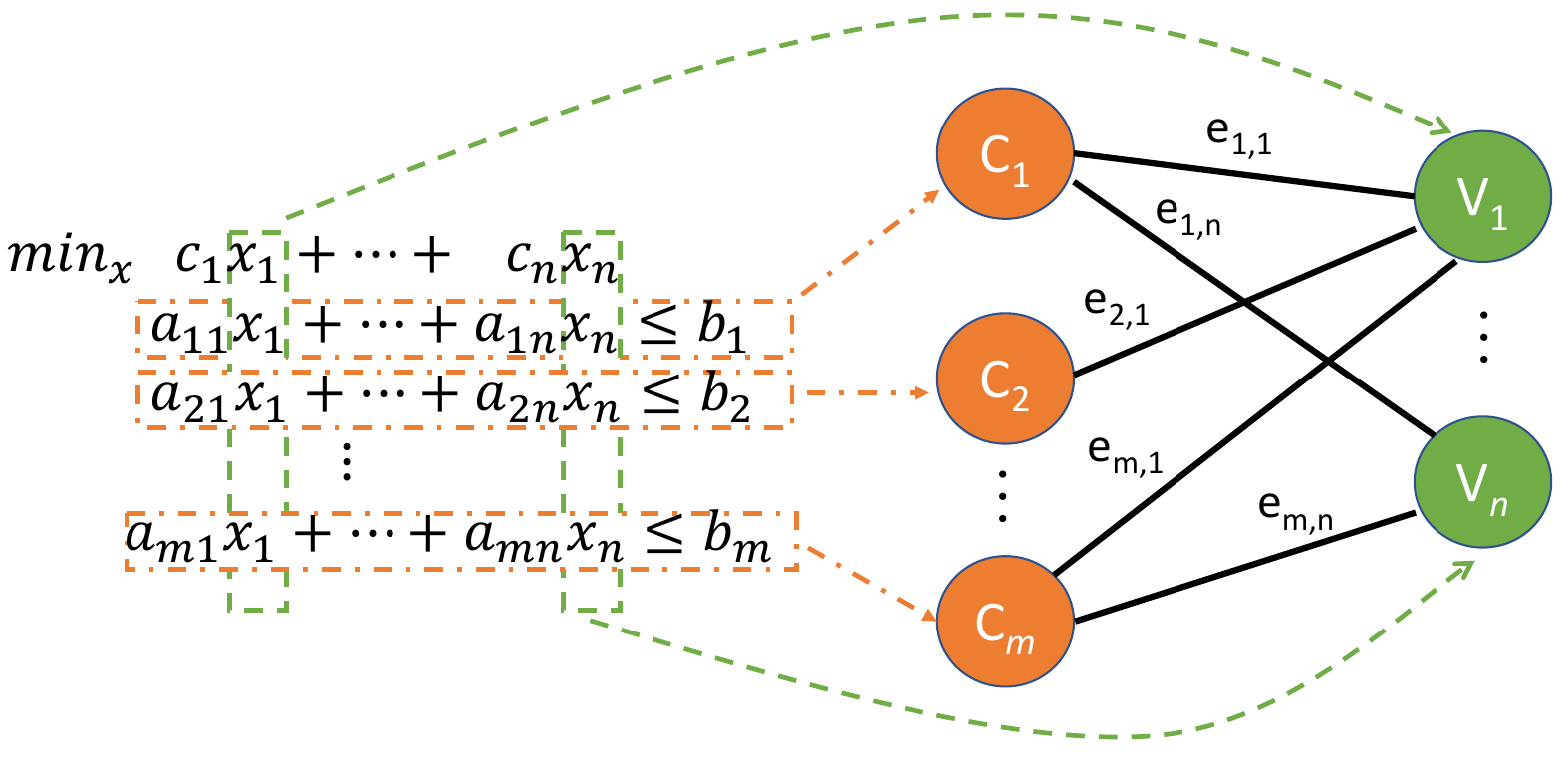}}
\caption{MIP model-based bipartite graph}
\label{bipartite graph}
\end{figure}

For any MIP, a bipartite graph can be defined as $\mathcal{G}=(\mathbf{C},\mathbf{E},\mathbf{V})$, where $\mathbf{C}$ is the set of constraint nodes, $\mathbf{E}$ is the set of edges, and $\mathbf{V}$ is the set of variable nodes. As shown in Fig. \ref{bipartite graph}, variable node $v_n$ is connected to constraint node $c_m$ if variable $x_n$ is involved in the $m$-th constraint.

The features of the constraint node, edge, and variable node are defined in Table \ref{table:bipartite graph feature} in Appendix \ref{Appendix A}. 

\subsection{Network Architecture}
\begin{figure}[htbp]
\centerline{\includegraphics[width=7.5cm]{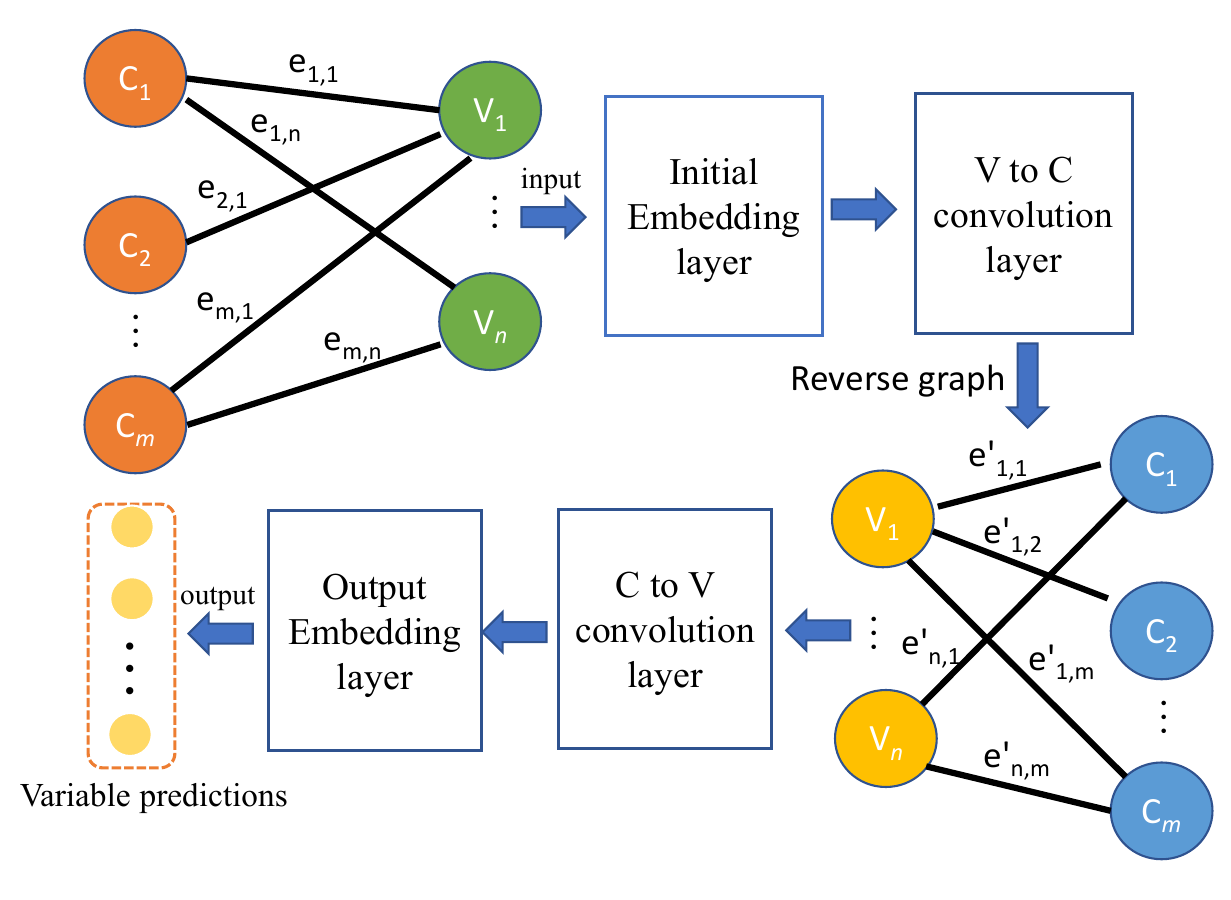}}
\caption{Network architecture of MB-GCN}
\label{bipartite network}
\end{figure}

The architecture of the baseline MB-GCN is shown in Fig. \ref{bipartite network}. Initially, the bipartite graph is passed through an embedding layer, followed by its traversal through a graph convolution layer. This layer facilitates the transmission of information from variable nodes to constraint nodes. Subsequently, the graph is reversed and directed to another graph convolution layer, enabling the propagation of information from constraint nodes back to variable nodes. Finally, the graph is fed to an output embedding layer, generating the output that signifies variable predictions.

The two graph convolution layers in Fig. \ref{bipartite network} have the same network structure but are initialized with different weights. For the V to C convolution layer, nodes are propagated using the operator in (\ref{eq:v2c convolution}):
\begin{equation} \label{eq:v2c convolution}
    x'_i = \mathbf{W}_C x_i + \sum_{j \in \mathcal{N}(i)}\mathbf{W}_V x_j + \sum_{j \in \mathcal{N}(i)}\mathbf{W}_E e_{i,j},
\end{equation}
where $\mathbf{W}_C,\mathbf{W}_V, \text{and} \ \mathbf{W}_E$ are weight matrices. $\mathcal{N}(i)$ is the set of indices of the neighbor nodes of node $i$.

\subsection{Loss Function for Neural Diving and Branching}
The MB-GCN is applicable to both neural diving and neural branching. In the context of neural diving, the loss function remains consistent with our proposed model and is computed using (\ref{eq:diving loss}) to (\ref{eq:pro}).

For neural branching, the output of the graph neural network is a vector representing the parameterized SB score of integer variables. The objective is to minimize the loss function defined by the cross-entropy function in (\ref{eq:cross-entro}):
\begin{equation}\label{eq:cross-entro}
    L(\theta)=-\log(\frac{\exp{(\boldsymbol{y}_c)}}{\sum_{i\in \mathbb{V}_I} \exp{(\boldsymbol{y}_i)}}),
\end{equation}
where $\boldsymbol{y}$ is the output vector of the graph convolution network, $c$ is the index of the candidate variable selected by strong branching policy, and $\mathbb{V}_I$ is the set of binary variables. 

\section{Numerical Studies}
In this section, we assess the performance of PI-GCN and MB-GCN in the context of neural diving. Subsequently, we evaluate MB-GCN specifically in relation to neural branching. Following these individual assessments, we integrate both neural diving and neural branching into SCIP for a comprehensive joint evaluation. Lastly, we present a detailed analysis of the computing resource requirements for both models.
\subsection{Experimental and Algorithm Setup}
In this subsection, we give the experimental and algorithm setups. All algorithms were executed on a server with a 32-core AMD Ryzen Threadripper 3970X 3.7GHz CPU and three 10 Gigabit NVIDIA GeForce RTX 2080 Ti. We used the non-commercial SCIP Optimization Suite 6.0 \cite{GleixnerEtal2018OO} as the MILP solver, which is open-source and allows us to build our own branching policy. 

We applied neural diving and neural branching techniques to solve a 24-hour period UC problem of the IEEE 1354-bus system. The parameters of the IEEE 1354-bus system were adopted from an Open-Source Julia/JuMP Optimization Package for the Security-Constrained Unit Commitment (SCUC) Problem \cite{xavier2022unitcommitment}. The historical load data of the California Independent System Operator (CASIO) \cite{CASIO} from July 1, 2017, to October 22, 2020, was adopted and scaled to be suitable for the testing system. Then, we used 1000 days for generating training data, 100 days for generating validation data, and 100 days for testing.

For neural diving, we used Gurobi to solve training and validation instances and collect the best feasible solutions in the solving process. Note that here we use Gurobi rather than SCIP, for gathering training samples due to its superior ability to produce better solutions within the specified time limit. Specifically, we configured Gurobi with a 5-hour time limit to maximize the quality of the obtained solutions. Finally, we collected 1,000 samples for training, 100 samples for validation, and 100 samples for testing. For neural branching, we used SCIP and the strong branching policy to collect 10,000 training samples and 1,000 validation samples. The hyperparameters of the baseline and proposed graph convolution model are given in Table \ref{table:neural diving} and \ref{table:neural branching} in Appendix \ref{Appendix B}, which are tuned separately to reach their best performance. 

\subsection{Performance Comparison on Neural Diving}
In this subsection, we compare the performance of our proposed PI-GCN with the baseline MB-GCN. We randomly chose 10 days over 100 testing days to generate instances. We set the computing time limit of SCIP to 10 minutes. 

\begin{figure}[htbp]    
  \centering         
  \captionsetup[subfloat]{labelfont=footnotesize,textfont=footnotesize}
  \subfloat[validation loss]
  {
      \label{fig:subfig1}\includegraphics[width=0.32\textwidth]{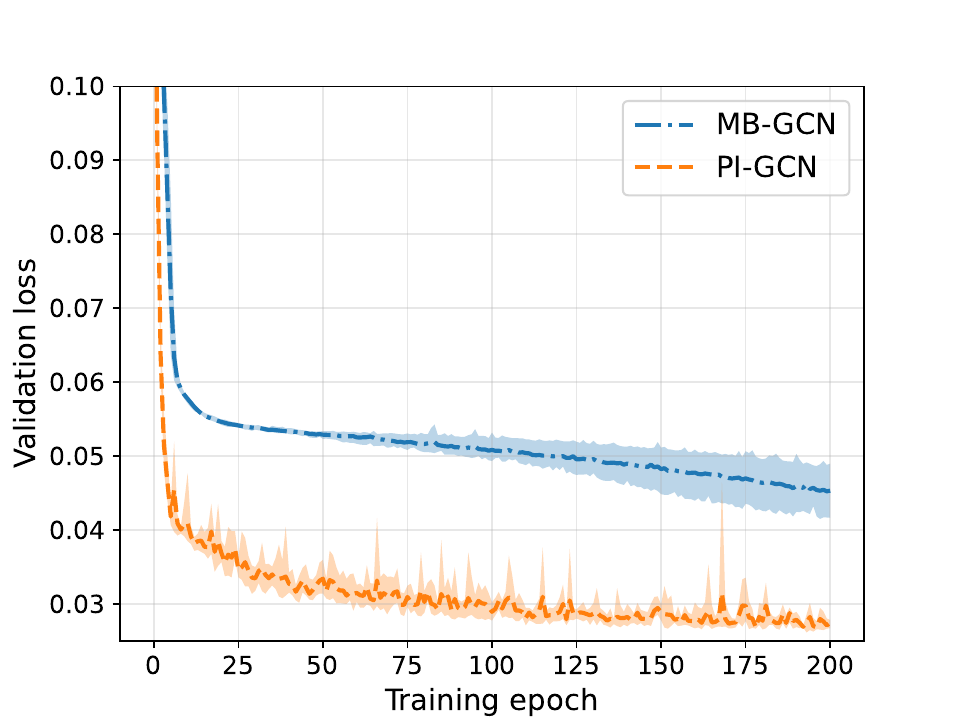}
  } 
  \\
  \subfloat[validation accuracy]
  {
      \label{fig:subfig2}\includegraphics[width=0.32\textwidth]{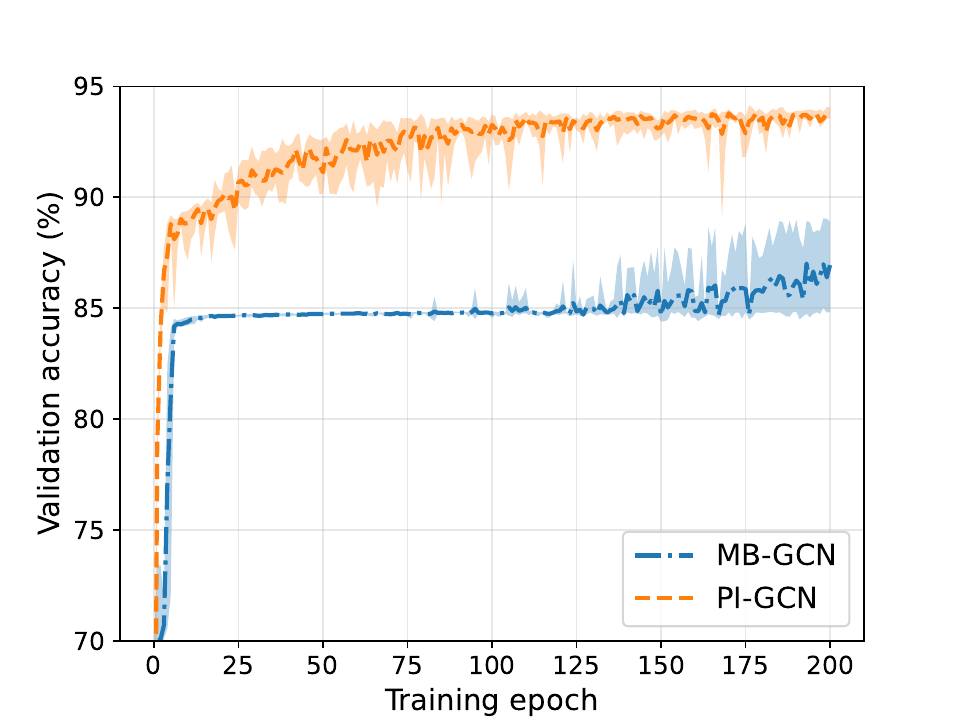}
  }
  \caption{Validation loss and accuracy of neural diving models}
  \label{fig:valid_loss_acc}            
\end{figure}

Figure \ref{fig:valid_loss_acc} illustrates the validation losses and accuracy of both the baseline MB-GCN model and our proposed PI-GCN model. The dark-colored lines depict the average performance across 5 random experiments, while the light-colored regions indicate the associated error bounds. Notably, the validation loss of PI-GCN converges to a consistently lower level compared to MB-GCN during the training process. Additionally, PI-GCN demonstrates superior prediction accuracy of binary variables when compared to MB-GCN.

After completing the training processes, we begin the testing phase by employing both trained models on the testing dataset to assess their predictive capabilities. Throughout the entire testing phase, the parameters that yield the highest prediction accuracy on the validation dataset are utilized for both models. The accuracy values presented in Table \ref{tab:testing_acc} are computed using (\ref{eq:acc}). The data in Table \ref{tab:testing_acc} reveals that, within each accuracy interval, the proposed PI-GCN model possesses a greater quantity of binary variables compared to MB-GCN.

\begin{table}[htbp]
    \caption{Number of binary variables in different accuracy intervals}
    \begin{center}
    \begin{tabular}{c|c|c|c|c|c}
    \hline
       Model & $\geq 80\%$  & $\geq 85\%$ & $\geq 90\%$ & $\geq 95\%$ & $= 100\%$\\ \hline
       MB-GCN & 4,582 &4,431 &4,253 &4,117 &3,554 \\ \hline
       PI-GCN &4,846 &4,708 &4,533 &4,349 &3,692 \\ \hline
    \end{tabular}
    \end{center}
    \label{tab:testing_acc}
    \begin{tablenotes}
        \item Note: The total number of binary variables to be predicted is 6,240 for the 1354-bus system.
    \end{tablenotes}
\vspace{-0.3cm}
\end{table}

Next, we apply both neural diving models to solve testing instances following Algorithm \ref{algo:parallel}. Here we set the accuracy threshold $\epsilon=0.95$ and select a sample ratio set $\mathcal{P}=[0.75,1.00]$ with intervals of 0.05. We also give the result of plain SCIP to show the improvement of both models. The default branching rule used by SCIP is reliability branching on pseudo cost values (relpcost). For convenience, we name the plain SCIP as SCIP-relpcost. 
\vspace{-0.3cm}
\begin{table}[htbp]
    \caption{Operational cost of testing instances (k$\$$)}
    \centering
    \begin{tabular}{c|c|c|c}
    \hline
       Instance & \thead{SCIP-relpcost}  & \thead{SCIP-relpcost \\with MB-GCN} & \thead{SCIP-relpcost \\with PI-GCN}\\ \hline
       1 &19,991 &20,014 &19,919 \\ \hline
       2 &24,245 &24,186 &24,112 \\ \hline
       3 &24,328 &24,280 &24,218 \\ \hline
       4 &23,766 &23,777 &23,717 \\ \hline
       5 &23,727 &23,648 &23,578 \\ \hline
       6 &24,338 &24,281 &24,176 \\ \hline
       7 &24,302 &24,236 &24,175 \\ \hline
       8 &20,302 &20,240 &20,180 \\ \hline
       9 &24,068 &24,008 &23,936 \\ \hline
       10 &17,505 &17,533 &17,479 \\ \hline
    \end{tabular}
    \label{tab:dive_obj}
\end{table}

As depicted in Table \ref{tab:dive_obj}, our proposed PI-GCN for neural diving attains the lowest operational costs across all testing instances. On the other hand, in specific instances, neural diving with MB-GCN even surpasses SCIP-relpcost's costs due to erroneous binary variable predictions.

\begin{figure}[htbp]
\centerline{\includegraphics[width=7.0cm]{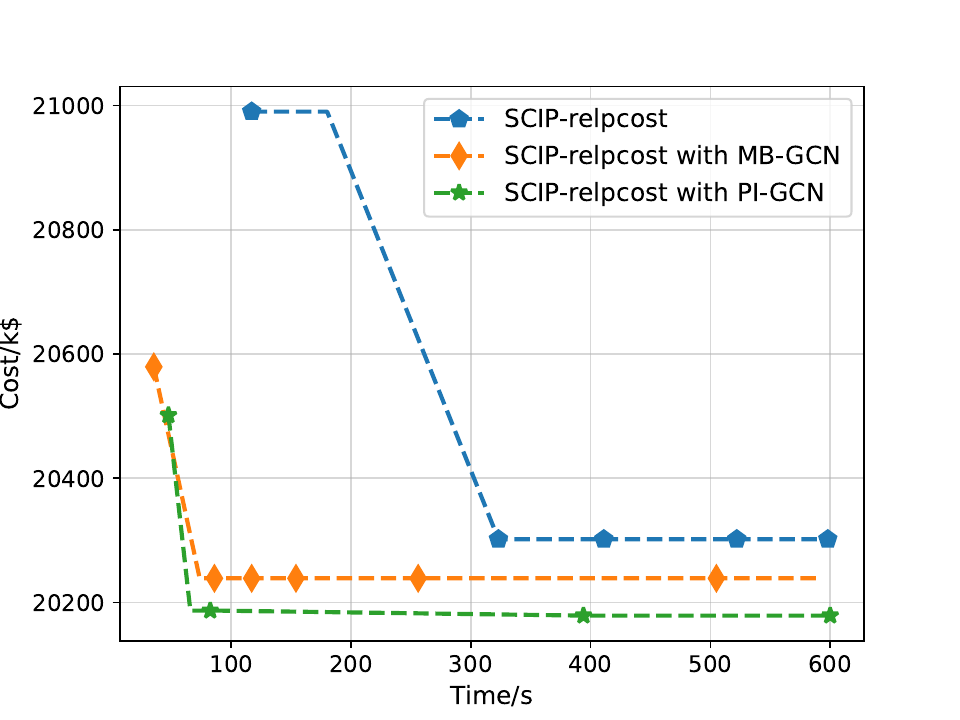}}
\caption{Operational cost versus solving time of instance 8}
\label{fig:gap_time_dive}
\end{figure}

To further demonstrate the improvement of our proposed model, the obtained operational costs with respect to computation time reported by SCIP during the solving process, are depicted in Fig. \ref{fig:gap_time_dive}. Due to space limitations, only the results of instance 8 are reported. We can see that SCIP-relpcost with MB-GCN and SCIP-relpcost with PI-GCN first reach feasible solutions within 50 s while SCIP-relpcost reaches a feasible solution around 110 s. Besides, SCIP with PI-GCN provides the lowest operational cost (20,180 k\$) within 10 minutes which is lower than SCIP-relpcost with MB-GCN (20,240 k\$) and SCIP-relpcost (20,302 k\$).

\vspace{-0.5cm}
\subsection{Performance of Neural Branching Technique}\label{sec:neural_branch}
In this subsection, we evaluate the performance of MB-GCN for neural branching. Due to space constraints, we forgo the training process. Following the completion of the training, we utilize the trained model to solve testing instances within a 10-minute time limit.

As shown in Table \ref{tab:neural_branch}, using neural branching (SCIP-neurb) as the custom branch policy achieves a lower MIP gap than using reliability branching on pseudo cost values. The main reason for the low performance of neural branching in some instances is the use of the three-binary model, which results in a hard MIP problem with 18980 binary variables for the IEEE 1354-bus system. Moreover, MB-GCN relies heavily on computing resources, leading to sub-optimal performance.


\begin{table}[htbp]
    \caption{MIP gap of testing instances ($\%$)}
    \centering
    \begin{tabular}{c|c|c}
    \hline
       Instance & \thead{SCIP-relpcost}  & \thead{SCIP-neurb}\\ \hline
       1 & 3.93 &3.54 \\ \hline
       2 &3.83  &3.41 \\ \hline
       3 &3.13  &2.39  \\ \hline
       4 &3.62  &2.94  \\ \hline
       5 &3.45 &2.64  \\ \hline
       6 &2.53 &2.17  \\ \hline
       7 &3.18 &2.78  \\ \hline
       8 &3.23 &2.68 \\ \hline
       9 &2.16 &1.75 \\ \hline
       10 &2.31 &1.91  \\ \hline
    \end{tabular}
    \label{tab:neural_branch}
\end{table}

\begin{figure}[htbp]
\centerline{\includegraphics[width=6.5cm]{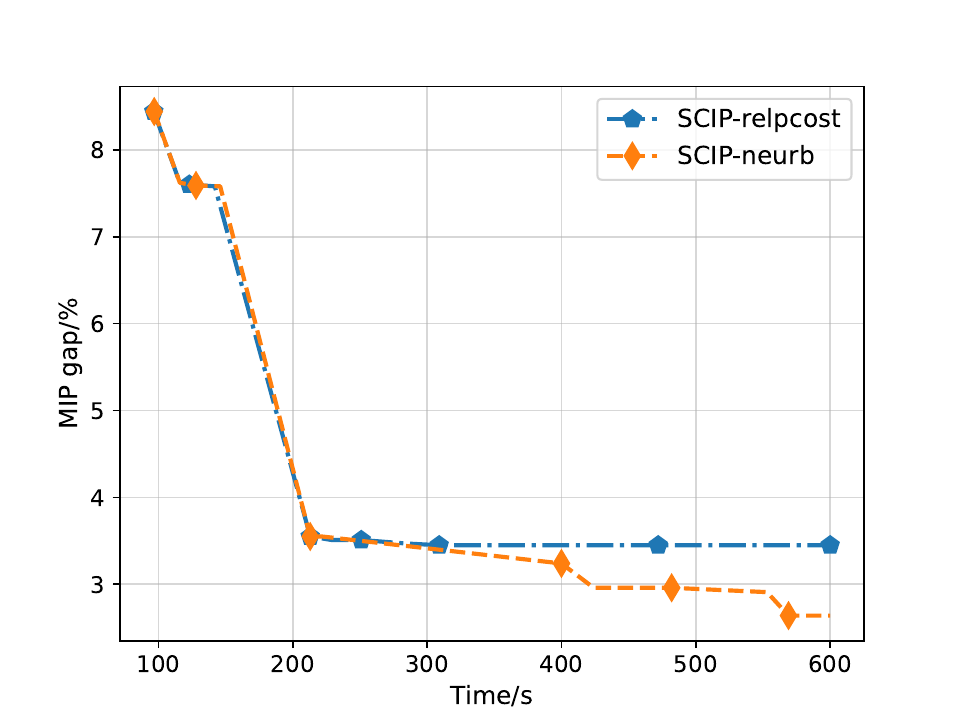}}
\caption{MIP gap versus solving time of instance 8}
\label{fig:gap_time_branch}
\end{figure}

Similarly, we show the MIP gap in relation to the computation time reported by SCIP of instance 8 in Fig. \ref{fig:gap_time_branch}. We can see that, after 200 s, the MIP gap obtained by reliability branching on pseudo cost values stays at 3.45\% while the MIP gap obtained by neural branching keeps decreasing to 2.64\%.

\vspace{-0.5cm}
\subsection{Joint Evaluation}
In this subsection, we integrate neural diving and neural branching into SCIP to do a joint evaluation. Specifically, we replace the default branch policy in SCIP with the neural branching model trained in subsection \ref{sec:neural_branch} and subsequently employ neural diving to solve the testing instances according to Algorithm \ref{algo:parallel}. 

The operational costs of testing instances are depicted in Table \ref{tab:neural_branch_dive} where we can see that SCIP-neurb with PI-GCN yields the smallest operational cost for all testing instances. Upon introducing neural branching, the baseline neural diving model is able to attain a lower cost in most of the testing instances, although, in certain instances, it yields an even higher operational cost. By contrast, the cost of all testing days is further reduced after combining our proposed neural diving model with neural branching. 

\begin{table}[htbp]
    \caption{Operational cost of testing instances (k$\$$)}
    \centering
    \begin{tabular}{c|c|c|c}
    \hline
       Instance & \thead{SCIP-neurb}  & \thead{SCIP-neurb \\with MB-GCN} & \thead{SCIP-neurb \\with PI-GCN}\\ \hline
       1 &19,916 &19,938 &19,844 \\ \hline
       2 &24,147 &24,088 &24,015 \\ \hline
       3 &24,153 &24,105 &24,043 \\ \hline
       4 &23,610 &23,621 &23,561 \\ \hline
       5 &23,540 &23,462 &23,393 \\ \hline
       6 &24,251 &24,194 &24,090 \\ \hline
       7 &24,209 &24,143 &24,082 \\ \hline
       8 &20,194 &20,131 &20,072 \\ \hline
       9 &23,972 &23,913 &23,841 \\ \hline
       10 &17,437 &17,464 &17,410 \\ \hline
    \end{tabular}
    \label{tab:neural_branch_dive}
\end{table}

\begin{figure}[htbp]
\centerline{\includegraphics[width=7.5cm]{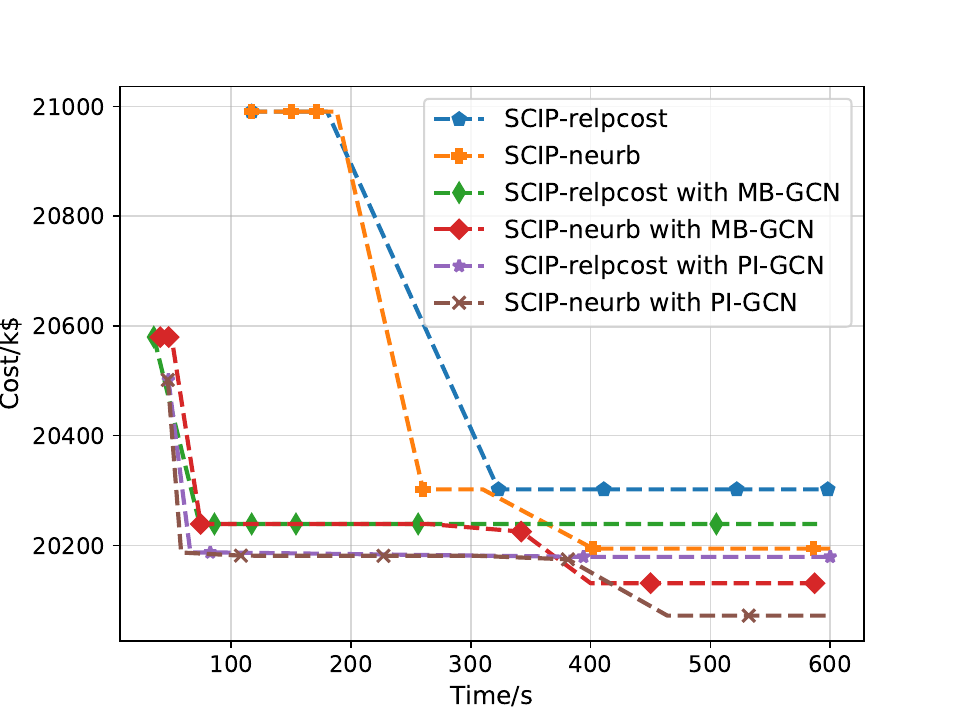}}
\caption{Comprehensive performance comparison of instance 8}
\label{fig:obj_time_dive_branch}
\end{figure}

Upon evaluating the performance of all algorithms for instance 8, as illustrated in Fig. \ref{fig:obj_time_dive_branch}, it is evident that both the baseline model and the proposed model for neural diving have lower operational costs when combined with neural branching. Finally, SCIP-neurb with PI-GCN produces the best result, with the lowest cost of 20,072 k\$. These results highlight the effectiveness of integrating neural branching with neural diving to achieve improved computational efficiency.

\subsection{Computing Resource Requirement and Scalability}
In this subsection, we provide a detailed analysis of the computing resource requirements for both MB-GCN and PI-GCN on various power systems to evaluate their scalability. Specifically, we report the maximum number of nodes and edges of the graphs generated by both models as well as the maximum allocated GPU memory of each model. To ensure a fair comparison of GPU memory usage, we set the epoch size and batch size to one and performed backpropagation for each batch. Moreover, we use PyTorch to accurately report the maximum amount of GPU memory used during training. Note that for all systems, we use the same hyperparameters of neural networks for both models, which are displayed in Table \ref{table:neural diving} to Table \ref{table:neural branching} in Appendix \ref{Appendix B}. 

As shown in Table \ref{tab:GPU requirement}, our proposed PI-GCN exhibits a significant reduction in the maximum node and edge inputs, with several orders of magnitude less than those of MB-GCN. Additionally, PI-GCN requires substantially less GPU memory allocation than MB-GCN, indicating its superior scalability. These findings provide compelling evidence of the scalability and versatility of our proposed model, and its ability to operate effectively across a range of systems with different scales. As the maximum allocated memory of MB-GCN was larger than our server's GPU capacity for IEEE 1888-bus, 2383-bus, and 3012-bus systems, we resorted to obtaining the maximum allocated memories of MB-GCN with various embedding sizes and using a curve-fitting method to estimate the maximum allocated memory for the given embedding size.

\begin{table}[htbp]
\begin{threeparttable}
\caption{GPU memory requirement of different systems}
\begin{tabular}{c|c|c|c}
\hline
\thead{System} &\thead{Max node input} &\thead{Max edge input} &\thead{Max allocated \\GPU memory(GB)}  \\
\hline
1354-bus	&137,621 (1,354) 	&7,452,282 (1,991) 	&7.04 (0.36) \\
\hline
1888-bus	&155,303 (1,888)	&13,030,995 (2,531)    &12.36 (0.57) 	  \\
\hline
2383-bus	&209,535 (2,383) 	&42,758,766 (2,896) 	&40.02 (1.21)  	\\
\hline
3012-bus	&216,110 (3,012) 	&52,104,887 (3,572)	&48.77 (1.52) \\
\hline
\end{tabular}
\label{tab:GPU requirement}
\begin{tablenotes}
  \item Note: The values inside and outside the parentheses represent data from PI-GCN and MB-GCN, respectively.
\end{tablenotes}
\end{threeparttable}
\end{table}

\section{Conclusion}
In this paper, we present a novel neural MIP solver that consists of neural diving and neural branching to solve large-scale UC problems. A physics-informed convolutional network (PI-GCN) is introduced for neural diving, which is trained to find high-quality variable assignments. To leverage the distinctive features of various components of power systems, we construct a spatiotemporal graph using time series load data and a spatial graph using the information of generators and transmission lines. A hierarchical graph convolution model is developed to handle diverse graph types and map physical input to the variable output. For neural branching, we adopt an MIP model-based graph convolutional network (MB-GCN), which selects the most suitable variables for branching at each node of the B\&B tree. Finally, we integrate neural diving and neural branching into a modern MIP solver to jointly evaluate the performance. Numerical studies show that our proposed PI-GCN outperforms the baseline MB-GCN in achieving a low operational cost. Moreover, the scalability of our model is significantly enhanced by reducing graph size and the dependence on computing resources. After combining with neural branching, our proposed neural diving model achieves the lowest operational cost for all testing days, demonstrating the effectiveness of the neural MIP solver in tackling large-scale UC problems.


 
\bibliographystyle{IEEEtran}
\bibliography{ref.bib}

\begin{thebibliography}{10}
\providecommand{\url}[1]{#1}
\csname url@samestyle\endcsname
\providecommand{\newblock}{\relax}
\providecommand{\bibinfo}[2]{#2}
\providecommand{\BIBentrySTDinterwordspacing}{\spaceskip=0pt\relax}
\providecommand{\BIBentryALTinterwordstretchfactor}{4}
\providecommand{\BIBentryALTinterwordspacing}{\spaceskip=\fontdimen2\font plus
\BIBentryALTinterwordstretchfactor\fontdimen3\font minus
  \fontdimen4\font\relax}
\providecommand{\BIBforeignlanguage}[2]{{%
\expandafter\ifx\csname l@#1\endcsname\relax
\typeout{** WARNING: IEEEtran.bst: No hyphenation pattern has been}%
\typeout{** loaded for the language `#1'. Using the pattern for}%
\typeout{** the default language instead.}%
\else
\language=\csname l@#1\endcsname
\fi
#2}}
\providecommand{\BIBdecl}{\relax}
\BIBdecl

\bibitem{gao2022internally}
Q.~Gao, Z.~Yang, W.~Yin, W.~Li, and J.~Yu, ``Internally induced branch-and-cut
  acceleration for unit commitment based on improvement of upper bound,''
  \emph{IEEE Trans. Power Syst.}, vol.~37, no.~3, pp. 2455--2458, 2022.

\bibitem{ongsakul2004unit}
W.~Ongsakul and N.~Petcharaks, ``Unit commitment by enhanced adaptive
  {Lagrangian} relaxation,'' \emph{IEEE Trans. Power Syst.}, vol.~19, no.~1,
  pp. 620--628, 2004.

\bibitem{hua2017representing}
B.~Hua, R.~Baldick, and J.~Wang, ``Representing operational flexibility in
  generation expansion planning through convex relaxation of unit commitment,''
  \emph{IEEE Trans. Power Syst.}, vol.~33, no.~2, pp. 2272--2281, 2017.

\bibitem{bertsimas2012adaptive}
D.~Bertsimas, E.~Litvinov, X.~A. Sun, J.~Zhao, and T.~Zheng, ``Adaptive robust
  optimization for the security constrained unit commitment problem,''
  \emph{IEEE Trans. Power Syst.}, vol.~28, no.~1, pp. 52--63, 2012.

\bibitem{ramesh2021accelerated}
A.~V. Ramesh, X.~Li, and K.~W. Hedman, ``An accelerated-decomposition approach
  for security-constrained unit commitment with corrective network
  reconfiguration,'' \emph{IEEE Trans. Power Syst.}, vol.~37, no.~2, pp.
  887--900, 2021.

\bibitem{yang2016multi}
L.~Yang, J.~Jian, Z.~Dong, and C.~Tang, ``Multi-cuts outer approximation method
  for unit commitment,'' \emph{IEEE Trans. Power Syst.}, vol.~32, no.~2, pp.
  1587--1588, 2016.

\bibitem{wu2013stochastic}
H.~Wu and M.~Shahidehpour, ``Stochastic {SCUC} solution with variable wind
  energy using constrained ordinal optimization,'' \emph{IEEE Trans. Sustain.
  Energy}, vol.~5, no.~2, pp. 379--388, 2013.

\bibitem{an2014exploring}
Y.~An and B.~Zeng, ``Exploring the modeling capacity of two-stage robust
  optimization: Variants of robust unit commitment model,'' \emph{IEEE Trans.
  Power Syst.}, vol.~30, no.~1, pp. 109--122, 2014.

\bibitem{gurobi}
\BIBentryALTinterwordspacing
{Gurobi Optimization, LLC}, ``{Gurobi Optimizer Reference Manual},'' 2023.
  [Online]. Available: \url{https://www.gurobi.com}
\BIBentrySTDinterwordspacing

\bibitem{cplex}
\BIBentryALTinterwordspacing
{IBM, LLC}, ``{ILOG CPLEX},'' 2023. [Online]. Available:
  \url{http://www.ilog.com/products/cplex}
\BIBentrySTDinterwordspacing

\bibitem{GleixnerEtal2018OO}
\BIBentryALTinterwordspacing
A.~Gleixner, M.~Bastubbe, L.~Eifler \emph{et~al.}, ``{The SCIP Optimization
  Suite 6.0},'' Optimization Online, Technical Report, July 2018. [Online].
  Available: \url{http://www.optimization-online.org/DB_HTML/2018/07/6692.html}
\BIBentrySTDinterwordspacing

\bibitem{FICO}
\BIBentryALTinterwordspacing
{Xpress optimization, LLC}, ``{Xpress optimization Reference Manual},'' 2023.
  [Online]. Available:
  \url{https://www.fico.com/en/products/fico-xpress-optimization}
\BIBentrySTDinterwordspacing

\bibitem{fu2013modeling}
Y.~Fu, Z.~Li, and L.~Wu, ``Modeling and solution of the large-scale
  security-constrained unit commitment,'' \emph{IEEE Trans. Power Syst.},
  vol.~28, no.~4, pp. 3524--3533, 2013.

\bibitem{morales2013tight}
G.~Morales-Espa{\~n}a, J.~M. Latorre, and A.~Ramos, ``Tight and compact {MILP}
  formulation for the thermal unit commitment problem,'' \emph{IEEE Trans.
  Power Syst.}, vol.~28, no.~4, pp. 4897--4908, 2013.

\bibitem{atakan2017state}
S.~Atakan, G.~Lulli, and S.~Sen, ``A state transition {MIP} formulation for the
  unit commitment problem,'' \emph{IEEE Trans. Power Syst.}, vol.~33, no.~1,
  pp. 736--748, 2017.

\bibitem{yan2019systematic}
B.~Yan, P.~B. Luh, T.~Zheng, D.~A. Schiro, M.~A. Bragin, F.~Zhao, J.~Zhao, and
  I.~Lelic, ``A systematic formulation tightening approach for unit commitment
  problems,'' \emph{IEEE Trans. Power Syst.}, vol.~35, no.~1, pp. 782--794,
  2019.

\bibitem{sun2018novel}
X.~Sun, P.~B. Luh, M.~A. Bragin, Y.~Chen, J.~Wan, and F.~Wang, ``A novel
  decomposition and coordination approach for large day-ahead unit commitment
  with combined cycle units,'' \emph{IEEE Trans. Power Syst.}, vol.~33, no.~5,
  pp. 5297--5308, 2018.

\bibitem{nikmehr2022quantum}
N.~Nikmehr, P.~Zhang, and M.~A. Bragin, ``Quantum distributed unit commitment:
  An application in microgrids,'' \emph{IEEE Trans. Power Syst.}, vol.~37,
  no.~5, pp. 3592--3603, 2022.

\bibitem{yang2021machine}
Y.~Yang and L.~Wu, ``Machine learning approaches to the unit commitment
  problem: Current trends, emerging challenges, and new strategies,'' \emph{The
  Electricity Journal}, vol.~34, no.~1, p. 106889, 2021.

\bibitem{chen2022security}
Y.~Chen, F.~Pan, F.~Qiu, A.~S. Xavier, T.~Zheng, M.~Marwali, B.~Knueven,
  Y.~Guan, P.~B. Luh, L.~Wu \emph{et~al.}, ``Security-constrained unit
  commitment for electricity market: Modeling, solution methods, and future
  challenges,'' \emph{IEEE Trans. Power Syst.}, vol.~38, no.~5, pp. 4668--4681,
  2022.

\bibitem{xavier2021learning}
{\'A}.~S. Xavier, F.~Qiu, and S.~Ahmed, ``Learning to solve large-scale
  security-constrained unit commitment problems,'' \emph{INFORMS Journal on
  Computing}, vol.~33, no.~2, pp. 739--756, 2021.

\bibitem{khalil2016learning}
E.~Khalil, P.~Le~Bodic, L.~Song, G.~Nemhauser, and B.~Dilkina, ``Learning to
  branch in mixed integer programming,'' in \emph{Proceedings of the AAAI
  Conference on Artificial Intelligence}, vol.~30, no.~1, 2016.

\bibitem{alvarez2017machine}
A.~M. Alvarez, Q.~Louveaux, and L.~Wehenkel, ``A machine learning-based
  approximation of strong branching,'' \emph{INFORMS Journal on Computing},
  vol.~29, no.~1, pp. 185--195, 2017.

\bibitem{marcos2015machine}
A.~Marcos~Alvarez, L.~Wehenkel, and Q.~Louveaux, ``Machine learning to balance
  the load in parallel branch-and-bound,'' \emph{Technical Report}, 2015.

\bibitem{khalil2017learning}
E.~B. Khalil, B.~Dilkina, G.~L. Nemhauser, S.~Ahmed, and Y.~Shao, ``Learning to
  run heuristics in tree search.'' in \emph{IJCAI}, 2017, pp. 659--666.

\bibitem{khalil2017alearning}
E.~Khalil, H.~Dai, Y.~Zhang, B.~Dilkina, and L.~Song, ``Learning combinatorial
  optimization algorithms over graphs,'' \emph{Advances in neural information
  processing systems}, vol.~30, 2017.

\bibitem{liao2021review}
W.~Liao, B.~Bak-Jensen, J.~R. Pillai, Y.~Wang, and Y.~Wang, ``A review of graph
  neural networks and their applications in power systems,'' \emph{Journal of
  Modern Power Systems and Clean Energy}, vol.~10, no.~2, pp. 345--360, 2021.

\bibitem{gasse2019exact}
M.~Gasse, D.~Ch{\'e}telat, N.~Ferroni, L.~Charlin, and A.~Lodi, ``Exact
  combinatorial optimization with graph convolutional neural networks,''
  \emph{Advances in Neural Information Processing Systems}, vol.~32, 2019.

\bibitem{gupta2020hybrid}
P.~Gupta, M.~Gasse, E.~Khalil, P.~Mudigonda, A.~Lodi, and Y.~Bengio, ``Hybrid
  models for learning to branch,'' \emph{Advances in neural information
  processing systems}, vol.~33, pp. 18\,087--18\,097, 2020.

\bibitem{labassi2022learning}
A.~G. Labassi, D.~Ch{\'e}telat, and A.~Lodi, ``Learning to compare nodes in
  branch and bound with graph neural networks,'' \emph{arXiv preprint
  arXiv:2210.16934}, 2022.

\bibitem{nair2020solving}
V.~Nair, S.~Bartunov, F.~Gimeno, I.~von Glehn, P.~Lichocki, I.~Lobov,
  B.~O'Donoghue, N.~Sonnerat, C.~Tjandraatmadja, P.~Wang \emph{et~al.},
  ``Solving mixed integer programs using neural networks,'' \emph{arXiv
  preprint arXiv:2012.13349}, 2020.

\bibitem{knueven2020mixed}
B.~Knueven, J.~Ostrowski, and J.-P. Watson, ``On mixed-integer programming
  formulations for the unit commitment problem,'' \emph{INFORMS Journal on
  Computing}, vol.~32, no.~4, pp. 857--876, 2020.

\bibitem{berthold2006primal}
T.~Berthold, ``Primal heuristics for mixed integer programs,'' Ph.D.
  dissertation, Zuse Institute Berlin (ZIB), 2006.

\bibitem{eckstein2007pivot}
J.~Eckstein and M.~Nediak, ``Pivot, cut, and dive: a heuristic for 0-1 mixed
  integer programming,'' \emph{Journal of Heuristics}, vol.~13, no.~5, pp.
  471--503, 2007.

\bibitem{yu2017spatio}
B.~Yu, H.~Yin, and Z.~Zhu, ``Spatio-temporal graph convolutional networks: A
  deep learning framework for traffic forecasting,'' \emph{arXiv preprint
  arXiv:1709.04875}, 2017.

\bibitem{defferrard2016convolutional}
M.~Defferrard, X.~Bresson, and P.~Vandergheynst, ``Convolutional neural
  networks on graphs with fast localized spectral filtering,'' \emph{Advances
  in neural information processing systems}, vol.~29, 2016.

\bibitem{xavier2022unitcommitment}
A.~S. Xavier, A.~M. Kazachkov, O.~Yurdakul, and F.~Qiu, ``Unitcommitment. jl: A
  julia/jump optimization package for security-constrained unit commitment
  (version 0.3),'' 2022.

\bibitem{CASIO}
\BIBentryALTinterwordspacing
{CASIO}, ``{California ISO Demand Forecast website},'' 2022. [Online].
  Available: \url{http://oasis.caiso.com/mrioasis/logon.do}
\BIBentrySTDinterwordspacing

\end{thebibliography}

\newpage

 




\clearpage
\appendices
\section{Features of Bipartite Graph}
\label{Appendix A}

The features of constraint, edge, and variable node of the bipartite graph of the baseline model are given in Table \ref{table:bipartite graph feature}. 
\begin{table}[h]
	\caption{Features of different nodes of bipartite graph}
    \begin{center}
	\begin{tabular}{L{0.3cm}| L{1.3cm}| L{5.9cm}}
		\hline  
        \multirow{1}{*}{\textbf{Set}} &\textbf{Feature} &\textbf{Description} \\ \hline
		\multirow{5}{*}{$\mathbf{C}$} &obj\_cos\_sim &Cosine similarity with objective \\ 
		    & bias  &Bias value, normalized with constraint coefficients  \\ 
		    & is\_tight &Tightness indicator in LP solution \\
		    & dual\_sol\_val &Dual solution value, normalized \\
		    & age &LP age, normalized with the total number of LPs \\ \hline    
		\multirow{1}{*}{$\mathbf{E}$} & coef &Constraint coefficient, normalized per constraint\\ 
		    \hline        
        \multirow{13}{*}{$\mathbf{V}$} & type & Type (binary, integer, impl. integer, continuous) as a one-hot encoding. \\
            & coef &Objective coefficient, normalized \\
            & has\_lb &Lower bound indicator\\ 
            & has\_ub &Upper bound indicator\\ 
            & sol\_is\_at\_lb &Solution value equals lower bound \\
            & sol\_is\_at\_ub &Solution value equals upper bound \\
            & sol\_frac & Solution value fractionality \\
            & basis\_status &Simplex basis status (lower, basic, upper, zero) as a one-hot encoding \\
            & reduced\_cost &Reduced cost, normalized \\
            & age &LP age, normalized \\
            & sol\_val &Solution value \\
            & inc\_val &Value in incumbent \\
            & avg\_inc\_val &Average value in incumbents \\
		\hline
	\end{tabular}
    \end{center}
	\label{table:bipartite graph feature}
\end{table}

\section{Hyperparameters of baseline and proposed model}
\label{Appendix B}

The hyperparameters of baseline MB-GCN and proposed PI-GCN for neural diving and neural branching are given in Table \ref{table:neural diving} to Table IX.

\begin{table}[h]
	\caption{Hyperparameters of PI-GCN for neural diving}
    \begin{center}
	\begin{tabular}{c|c|c|c} 
		\hline
    \multirow{13}{*}{\begin{tabular}[c]{@{}c@{}}STGC\\ Module\end{tabular}}
		&\multirow{3}{*}{\begin{tabular}[c]{@{}c@{}}1st Temporal \\ Conv Layer\end{tabular}}  &in\_channel size  & 64 \\
        &      &out\_channel size  & 16 \\
	  & 	 &kernel size &3 \\  \cline{2-4}
        &\multirow{3}{*}{\begin{tabular}[c]{@{}c@{}}Chebyshev\\ Conv Layer\end{tabular}}  &in\_channel size  & 16 \\
        &       &out\_channel size  & 64 \\
        &       &kernel size &3 \\  \cline{2-4}
		&\multirow{3}{*}{\begin{tabular}[c]{@{}c@{}}2st Temporal\\ Conv Layer\end{tabular}}  &in\_channel size  & 64 \\
        &      &out\_channel size  & 16 \\
	  & 	 &kernel size &3 \\  \cline{2-4}
		&\multirow{4}{*}{\begin{tabular}[c]{@{}c@{}}Fully-connected\\ Layer\end{tabular}}  &in\_channel size  & 128 \\
        &   &hidden size  & 128 \\
        &   &out\_channel size  & 64 \\
	  &	  &activation function &ReLU \\   \cline{1-4}

    \multirow{7}{*}{\begin{tabular}[c]{@{}c@{}}ECGC\\ Module\end{tabular}}
		&\multirow{3}{*}{\begin{tabular}[c]{@{}c@{}}Edge-conditioned\\ Conv Layer\end{tabular}}  &in\_channel size & 64 \\
        &      &out\_channel size & 64 \\
	  & 	 &number of layers &2 \\  \cline{2-4}
        &\multirow{4}{*}{\begin{tabular}[c]{@{}c@{}}Fully-connected\\ Layer\end{tabular}}  &in\_channel size  & 64 \\
        &       &hidden size  & 128 \\
        &       &out\_channel size  & 128 \\
        &       &activation function &ReLu \\  \cline{1-4}
    
   \multirow{7}{*}{\begin{tabular}[c]{@{}c@{}}Variable Map\\ Module\end{tabular}}
    &\multirow{3}{*}{\begin{tabular}[c]{@{}c@{}}Chebyshev\\ Layer\end{tabular}}  &in\_channel size & 128 \\
  & 	 &out\_channel size &64 \\
  & 	 &kernel size &9 \\\cline{2-4}
    &\multirow{4}{*}{\begin{tabular}[c]{@{}c@{}}Fully-connected\\ Layer\end{tabular}}  &in\_channel size  &64 \\
    &       &hidden size &128 \\  
    &       &out\_channel size &73 \\ 
    &       &activation function &Sigmoid  \\
		\hline
	\end{tabular}
    \end{center}
    \label{table:neural diving}
\end{table}
\newpage

\begin{table}[htbp]
	\caption{Hyperparameters of MB-GCN for neural diving}
    \begin{center}
	\begin{tabular}{c|c|c} 
		\hline 
		\multirow{3}{*}{Initial Embedding Layer}  &hidden size  & 24 \\
              &embedding size  & 24 \\  
              &activation function &ReLu \\ \cline{1-3}
        \multirow{3}{*}{V to C Conv Layer}  &hidden size  & 24 \\
               &embedding size  & 24 \\ 
               &activation function &ReLu \\ \cline{1-3}
        \multirow{3}{*}{C to V Conv Layer}  &hidden size  & 24 \\
               &embedding size  & 24 \\ 
               &activation function &ReLu \\ \cline{1-3}
        \multirow{3}{*}{Output Embedding Layer}  &hidden size  & 24 \\
              &embedding size  & 24 \\ 
              &activation function &ReLu \\ \cline{1-3}

    \multirow{9}{*}{\begin{tabular}[c]{@{}c@{}}Training Parameters\\ (Shared with proposed model)\end{tabular}} & number of epochs & $1000$ \\
                    & pre-train sample size &28 \\
                    & pre-train batch size &1 \\
                    & training batch size &1 \\
                    & validation batch size &1 \\
                    & epoch size  &10\\ 
                    & learning rate &0.005\\ 
                    & patience &50 \\ 
                    & early stop &100 \\
		\hline
	\end{tabular}
	\label{table:neural diving2}
    \vspace{1cm}
	\caption{Hyperparameters of MB-GCN for neural branching}
	\begin{tabular}{c|c|c} 
		\hline 
		\multirow{3}{*}{Initial Embedding Layer}  &hidden size  & 24 \\
              &embedding size  & 24 \\  
              &activation function &ReLu \\ \cline{1-3}
        \multirow{3}{*}{V to C Conv Layer}  &hidden size  & 24 \\
               &embedding size  & 24 \\ 
               &activation function &ReLu \\ \cline{1-3}
        \multirow{3}{*}{C to V Conv Layer}  &hidden size  & 24 \\
               &embedding size  & 24 \\ 
               &activation function &ReLu \\ \cline{1-3}
        \multirow{3}{*}{Output Embedding Layer}  &hidden size  & 24 \\
              &embedding size  & 24 \\ 
              &activation function &ReLu \\ \cline{1-3}

        \multirow{9}{*}{Training parameters} & number of epochs & $1000$ \\
                        & pre-train sample size &100 \\
                        & pre-train batch size &1 \\
                        & training batch size &1 \\
                        & validation batch size &1 \\
                        & epoch size  &100\\ 
                        & learning rate &0.005\\ 
                        & patience &100 \\ 
                        & early stop &200 \\
		\hline
	\end{tabular}
    \end{center}
    \label{table:neural branching}
\end{table}

\end{document}